\def \vc #1{{\mbox{\boldmath $#1$}}}

\def\thetag{{\vc \theta}}

\def\deltag{{\vc \delta}}
\def\gammag{{\vc \gamma}}

\documentstyle[11pt,mrs2001,epsfig]{article}
\begin{document}
\title{TRACING MATTER  
       WITH WEAK LENSING SURVEYS}

\author{Y. MELLIER $^{1,2}$, L. van WAERBEKE$^{1,3}$}
\affil{$^1$IAP, 98 bis boulevard Arago; 75014 Paris, France.}
\affil{$^2$Observatoire de Paris, DEMIRM 61 avenue de l'Observatoire; 75014 Paris, France.}  
\affil{$^3$CITA, Mc Lennan Labs., 60 St George Street; Toronto, Ont. M5S
3H8 Canada.}

\begin{abstract}
Gravitational weak lensing maps the location of (dark) matter at all 
scales. The lens-induced distortion
  field traces gravity fields and can be used to reconstruct the 
 mass distribution in galaxies, groups, clusters of galaxies 
 or large-scale structures.  From these reconstructions, one 
can in principle recover where the matter is, what are 
   its properties  and how it is coupled with light and baryons.  In the 
  following, we review the present status of weak lensing analysis and  
 discuss  the most recent results regarding matter properties in the 
  universe.  
\end{abstract}

\section{Introduction}
Gravitational lensing by foreground structures induces  
image  deformation of all distant galaxies.  Spectacular 
cases caused by infinite magnification 
eventually produce giant arcs and multiple images
configuration,   but these are exceptional events:
in general the amplitude of the distortion is  
weak and can be detected only statistically over a large number
of galaxies. In this 'weak lensing' regime, a non-parametric mass
reconstruction provides projected mass maps of the
Universe.
\\
The distortion field is given by
a line-of-sight projection of the mass density contrast.
This distortion field is measured from the shape of 
the lensed galaxies which is characterized by 
the  surface brightness second moments $I\left(\thetag\right)$, 
  (see \cite{mel99}, \cite{bart01} and references therein):

\begin{equation}
M_{ij}= \displaystyle{
 {\int
I\left(\thetag\right) \theta_i \ \theta_j
 \ {\rm d}^2\theta \over \int I(\thetag) \ {\rm d}^2\theta }} \ .
\end{equation}
A galaxy with intrinsic ellipticity ${\bf e}$ is then measured with an
ellipticity ${\bf e}+{\deltag}$, where ${\deltag}$ is the distortion
of the galaxies, given by
\begin{equation}
\deltag =2 \gammag \ {(1-\kappa)
\over
(1-\kappa)^2+\vert \gammag \vert^2} \ =\left(\delta_1:{M_{11}-M_{22}
\over Tr(M)} \ ;  \ \delta_2:{ 2 M_{12} \over Tr(
M)} \right) \ .
\label{distor}
\end{equation}
$\kappa$ and $\gammag$ are respectively the gravitational convergence 
  and shear, which both depend on the second derivative of the 
projected gravitational potential, $\varphi$:
\begin{equation}
 \kappa(\thetag) = \frac{1}{2}\,(\varphi_{,11}+\varphi_{,22})  ; \ \  \ \ \ \ \
  \gamma_1(\thetag) =
  \frac{1}{2}\left(\varphi_{,11}-\varphi_{,22}\right) ; \ \ \ \ \ \ \
  \gamma_2(\thetag) = \varphi_{,12} = \varphi_{,21} \ .
\label{convshear}
\end{equation}
Equations (\ref{distor}) and (\ref{convshear}) characterize 
  the relation between  the projected mass density 
  and the ellipticity of lensed galaxies.  Hence, they   
  provide a recipe to recover the projected gravity field 
  from the distortion field. In the case of weak 
lensing, the galaxy ellipticity is a direct measure of the shear
($\kappa<<1$ and $\vert \gamma \vert <<1$, and therefore 
  $\delta \approx 2 \ \gamma$) which makes the mass reconstruction
a simple linear operation.
\\
During the past decade, spectacular theoretical and technical  
  developments in weak lensing  analyses lead to the production 
of a new mass estimator which in principle works at all scales.
  Weak lensing reconstruction has been mostly used on galaxies, 
clusters of galaxies and large-scale structures.  The scientific outcomes
coming from these different analysis are presented and discussed in this review.

\section{Properties of galaxy halos}

Diagnostics about the properties of galaxy halos are important 
tests of CDM scenarios.  According to the numerical simulations 
of cosmic structures, galaxies are very dense systems
with a cuspy mass profile and mass density slopes similar to 
NFW family. These predictions can be tested 
with galaxy-galaxy lensing or mass reconstruction 
using weak distortion maps or strong lensing statistics.
\\
The galaxy-galaxy lensing analysis consists in separating
a foreground population of galaxies ({\sl ie} the lenses) from a
background sample ({\sl ie} the lensed galaxies). An averaged distortion
is then measured around all the foreground galaxies.
The investigation of 
galaxy halos is done by comparing the expected 
distortion field, inferred from analytical mass profiles,
to the observed one, computed from ellipticities of background 
galaxies. The  properties of analytical profiles   
 are characterized by a velocity dispersion or 
circular velocity, a  physical scale, 
  and a slope.  
\\
The orientation $\varphi$ of the background galaxies with respect to the
line joining the galaxy and the
foreground lens is given by (\cite{brain96}):
\begin{equation}
p\left(\varphi\right)={2 \over \pi}\left[1-\gamma_t \left<{1 \over 
\epsilon^s}\right>\ {\rm cos}\left(2\varphi\right)\right].
\label{galgaleq}
\end{equation}
Which means that the lensed galaxies are preferencialy tangentially aligned with
the lens ($\varphi\simeq 90$).
$\epsilon^s$ is the ellipticity of the (unlensed) background galaxy
  and $\gamma_t$ is the tangential component of the gravitational shear.
In practice, we have to take into account the fact that mass profiles of nearby
foreground galaxies overlap (Schneider \& Rix 1997; \cite{rix97}), but it
does not change the general
principle of the technique which was pioneered by Tyson et al (1984:
\cite{tys94}).  
\\
Quantitative results regarding the properties of galaxy halos have been
first obtained by Brainerd et al (\cite{brain96}) and are still in progress 
   using big imaging surveys (see \cite{mel99}, 
 \cite{bart01}). 
  The most recent ones use jointly imaging and spectroscopic data in order
 to scale the amount of lensing mass and the total luminosity of
 galaxies from the redshift distribution 
  of foreground and background galaxies.  McKay et al 2001
  (\cite{mckay01}) and 
  Smith et al 2000 (\cite{smith00}) used respectively the SDSS and
 the LCRS survey and obtained
detailed galaxy-galaxy analysis.  Both agree with earlier
   works that typical velocity dispersion or circular velocity of
 lensing galaxies range  between 120 to 220 km/sec. The 
  radial distribution of the shear is compatible with a power law
close to an isothermal profile, but there is not yet  conclusive 
evidence they are better fit than NFW profile.  Likewise, the 
typical physical scale is still uncertain and ranges between 20 
to 250 $h^{-1}$ kpc. The upper limit seems compatible with the most 
   recent analyses and implies that the 
  contribution of galaxy halos to $\Omega_m$ could be very large, 
 possibly $0.1 < \Omega_{gal} < 0.25$. Wilson et al. (2001;
\cite{wils01}) 
adopted a different strategy to analyze early type galaxy halos
of their survey (they compared a projected luminosity map 
with a mass map from weak lensing reconstruction) and they reached
similar conclusions regarding $\Omega_{gal}$. 
\\
The cuspy nature of galaxy halo is hard to test using weak lensing 
analysis since it cannot scales smaller than $100~{\rm pc}$.
However, such small scales are accessible  with strong lensing statistics
which measures the frequency of events like multiple images of
lensed quasars. For instance, highly peaked mass density profiles
produces a central image strongly demagnified, not visible in practice.
This lensing property of cuspy 
  profiles is an important 
  difference from soft core mass models and this was exploited by 
    Keeton (2001; \cite{kee01}) to explore the inner profile of the lenses in
the CLASS sample of radio galaxies. It turns out that 
  only few lenses do have visible odd image, whereas 
   about 30\% of the sample should show one, according 
  to  theoretical expectations from CDM models. If Keeton's 
interpretation is correct, galaxies are more cuspy than CDM halos. 
  Surprisingly,  the   fraction of gravitational lenses  of
the CLASS sample lead to an  opposite conclusion on scale
  larger than 10 kpc: CDM halos are too concentrated compared 
to galaxies.  This paradox is somewhat confusing but is 
 a  very interesting diagnostic regarding the description 
  of dark galaxy halos and its interpretation in the context of
  galaxy formation processes at very small scale. 
\\
Within the class of models he studied,
Keeton addressed many  issues and concluded that his
results are robust, and the associated uncertainties well understood.
Nevertheless, it is not yet obvious that his conclusions can be compared
with galaxy-galaxy lensing because the scales probed in the later are much
larger. Moreover, it must be keep in mind that Keeton's 
  star+halo model 
  may be only a partial description of galaxies.  On the other hand, 
 if the size of halos are as large as 
  250 $h^{-1}$ kpc, then  galaxy-galaxy lensing parameters may be 
 significantly contaminated  by other foreground galaxies or by additional 
  effects produced by groups or clusters  of galaxies. 

\section{Clusters of galaxies}

Gravitational growth of structure formation  
 predicts that clusters of galaxies 
  are dynamically young gravitational systems 
still in formation. Numerical simulations clearly show that 
  the mass distribution of clusters of galaxies, their 
  radial mass profile  as well as 
  the evolution of cluster abundance with redshift 
   strongly depend on cosmological models and can be used 
  in order to measure the cosmological parameters,  
like $\Omega_m$ or $\sigma_8$. Earlier work on
  cluster lensing revealed however that  these systems 
  are in general dynamically and thermodynamically complex 
  and hard to interpret from their baryonic content only
using simple models.
\\
Because gravitational lensing analysis only probes matter 
regardless the complexity of its internal physical properties, 
  it is an interesting alternative tool to standard
  virial and X-ray studies.
 From a  weak lensing technical point of view,
clusters of galaxies are well suited  systems
for mass reconstruction: their mass-density contrast is high
enough ($>100$) to produce significant
  gravitational distortion, and their angular scale
  ($\ge $ 10 arc-minutes) is much larger than the typical
angular distance between the lensed background galaxies.
Therefore mass reconstruction in cluster of galaxies is quite
easy and robust, and the angular resolution good enough
for scientific purpose (about $1$ arcmin).
\\
The present status of weak lensing mass reconstruction of 
  distant clusters of galaxies is listed in \cite{mel01}. 
  This sample only contains clusters with a redshift larger than 0.1 
  and does not include those analyzed 
  with the magnification bias (or depletion) technique.   
 Despite an important dispersion due to an
heterogeneous sample, some general trends emerge.
  The averaged  mass-to-light ratios from
  weak lensing ($WL$) is $\left(M/L\right)_{WL}\approx 400 \ h$ (with
a dispersion of $\pm 200 \ h$) and the typical
velocity dispersion is 1000 km/sec.
Assuming that all the mass is contained in clusters,
this corresponds to $\Omega_{m-WL} \approx 0.28$, which is
   in good agreement with   X-ray
   or virial analyses as well as strong lensing studies based 
  of giant arc reconstruction in clusters of galaxies.  
\\
Unfortunately, the differences between 
isothermal, power law or NFW
``universal'' models are still smaller than 
 the errors of the measured mass profile
 (see for example \cite{clo00}). The fact that
giant arcs statistic is in favor of highly peak mass profile is
not enough to distinguish between the different theoretical profiles.
In fact, in view of the present signal-to-noise ratio of mass maps and
  the large family of analytical mass profiles which are 
  proposed, this issue will demand important improvements in the shear 
  measurement and to stack together many cluster profiles.
   It is important to notice that the difficulty to separate all 
  the analytical mass profiles from weak lensing reconstruction 
is coming from random noise and not from intrinsic problems with mass 
   reconstruction (which has been extensively studied and tested). This
is well confirmed from the comparison between 
   weak lensing and X-ray analyses: for instance, the recent 
  comparison of  lensing studies in Abell 2390  with Chandra observations 
  (\cite{allen01}) 
  show a remarkable agreement of the mass profiles which confirm 
 {\sl a posteriori} that mass reconstruction are reliable.
\\
In contrast to the  rather good agreement observed between 
  weak lensing and X-ray,
   frequent discrepancies are reported on small scales 
 with strong lensing reconstructions.  Part of the discrepancy 
   is due to gravitational lensing and is 
  likely a projection effect of matter located anywhere along 
  the line of sight of clusters (\cite{reblin99}, \cite{metz99}).
  But on very small scale, inside 
    the innermost
   regions,  physical properties of clusters  are complex and 
 a simplistic 
   description of hot baryons is no longer valid.  \cite{allen98}
 observed that the discrepancy is only significant in 
  clusters without cooling flow of intra-cluster gas.
  Since cooling flows are only detected in the densest
  systems which are also the most dynamically evolved, 
  the tendency he reported  reveals that only 
  young clusters show a discrepancy. If so,  it is likely
  that their dynamics and thermodynamics models of these clusters    
  are oversimplified.
\begin{figure}[t]
\centerline{\psfig{file=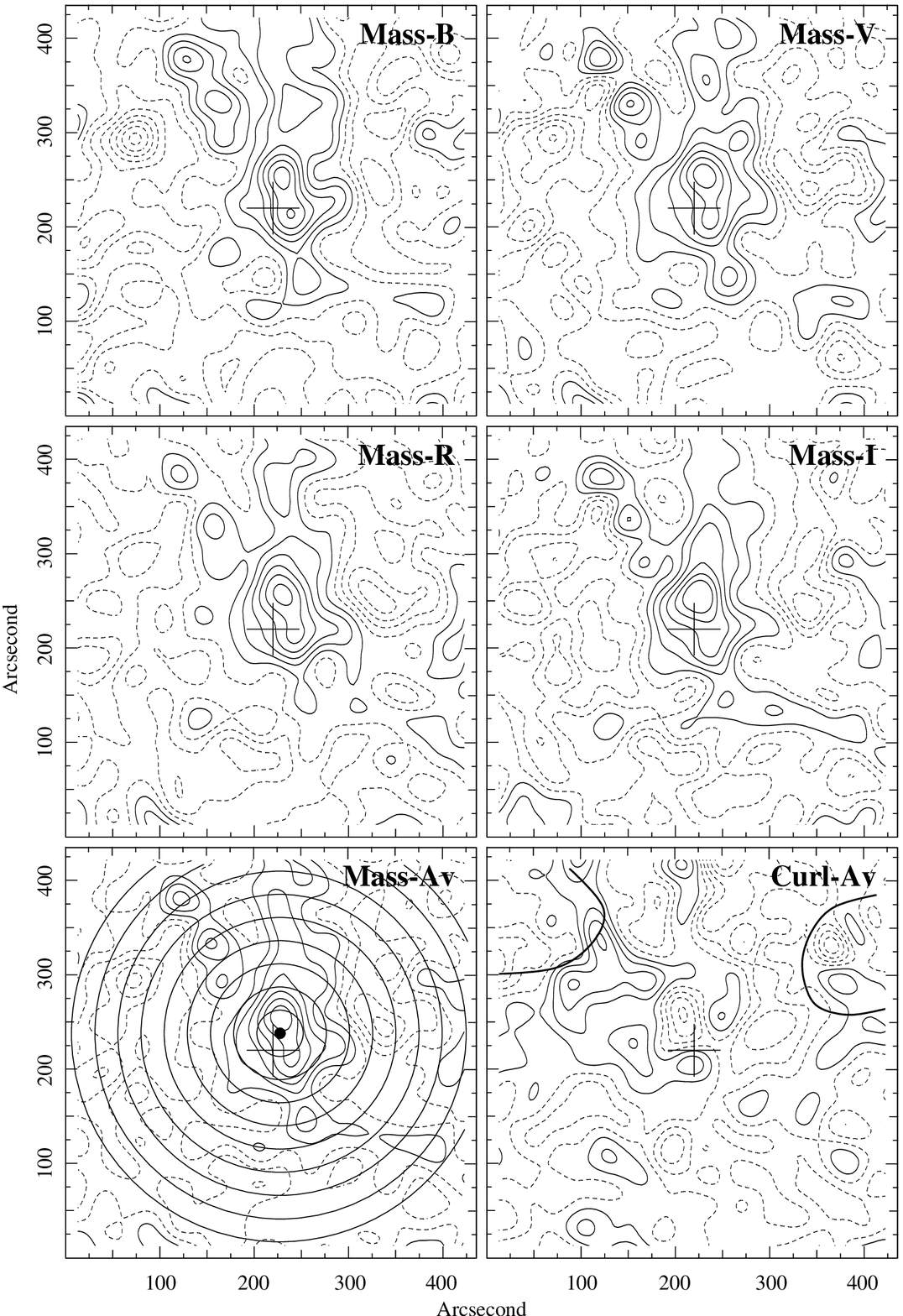,width=8.5cm} 
\psfig{file=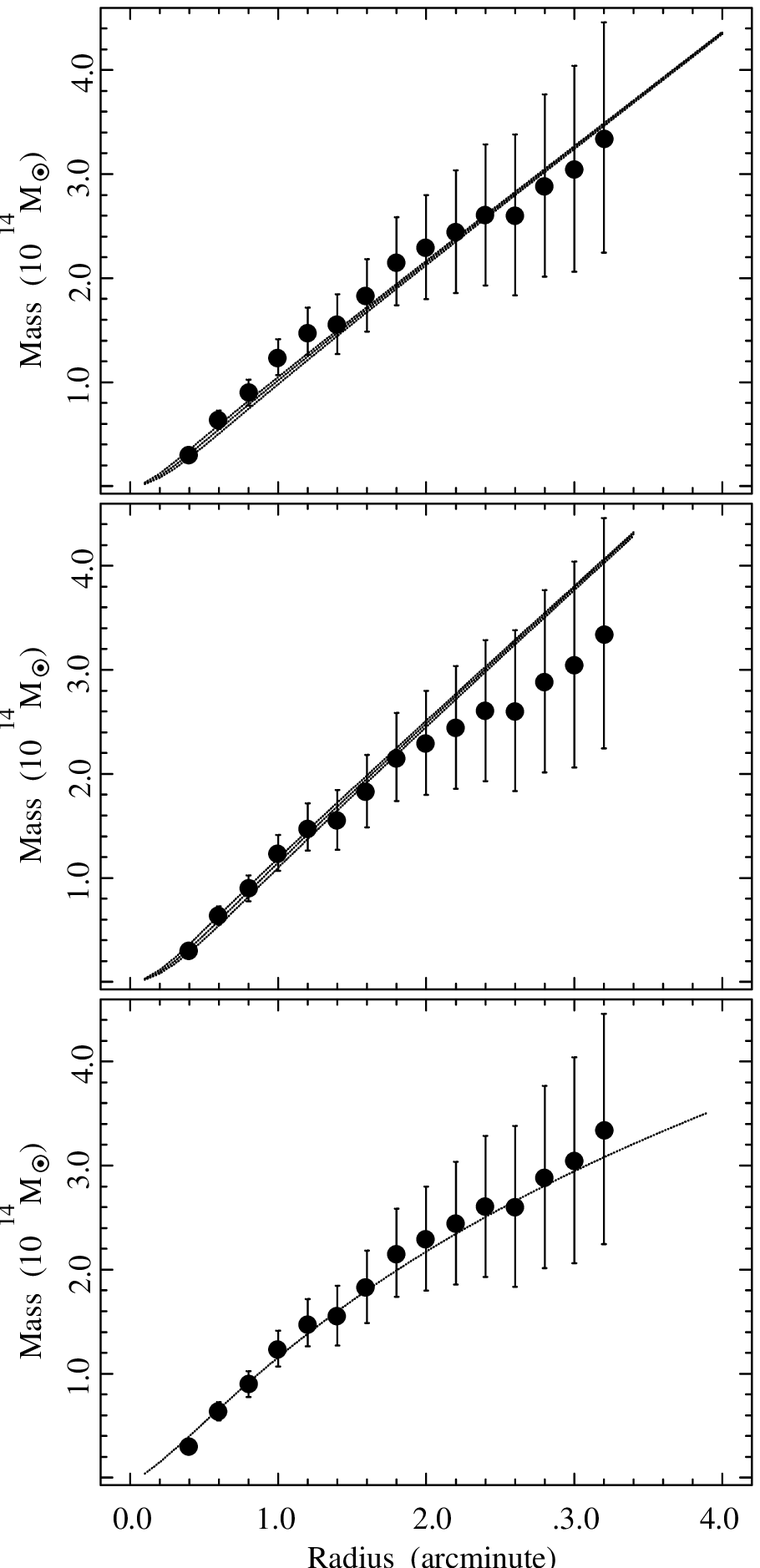,width=6cm} 
}
\caption{Mass reconstruction of the distant cluster MS1008 using weak 
lensing. The left panel show 4 mass reconstructions done independently
 using B,V, R and I data obtained at the VLT/UT1.  The bottom left panel show 
  the final
  reconstruction using the 4 filters jointly. The bottom right show
  random noise when galaxies are rotated by 90 degrees. On the right 
panel the radial mass profile is compared to an isothermal sphere 
  with 950 km/sec (top), 1080 km.sec (middle) and the best NFW fit
 (bottom).  Both top and bottom panel fit equally well the data
 (from \cite{athreya99}).
\label{ms1008}
}
\label{mass1008}
\end{figure}

\section{Dark clusters}
The unknown contribution to dark matter which could not be 
 detected by any method but gravitational 
lensing would be the so-called ``dark clusters''.   
   The most convincing cases reported so far (HST/STIS field,
 Miralles et al 2001 - private communication;
   Abell 1942, \cite{erben00}; Cl1604+4304,
  \cite{umet00}; Cl0024+1654, \cite{bon94})
  show a clear shear pattern spread over angular scale of one 
  arcminute.
   Their typical mass,  estimated from reasonable assumptions
  on their redshift, is about $10^{14}$ M$_{\odot}$
  and corresponds to $M/L >> 500$.
  \\
It is not clear yet whether dark clusters are real  physically bound
systems. An alternative view is that they are 
indeed very high redshift clusters: they are not detected 
 because they are too faint and because most of its optical light
  is shifted toward the infrared. This possibility has been explored 
 by \cite{gray01} who carried out very deep H-band 
  observations of Abell 1942.
   But nothing has been detected so far.  It is also 
possible that these systems do not have bright galaxies 
   but do have a hot intra-cluster gas.  So far, none of them 
  have been observed deeply in X-ray telescopes, so it is 
  not demonstrated yet that dark clusters do not contain 
  baryonic matter. 
\\
The key question is now to confirm that dark clusters exist 
or whether these detections are fortuitous fluctuations along
some line of sights. ``Dark clusters'' could also be produced by
intrinsic alignment, or just systematics. This is a difficult issue: up to now 
   we have no way, but gravitational lensing, to ``see'' them, 
  and if they contain neither galaxies nor hot gas, we cannot 
measure their redshift. From a cosmological point of view, the existence 
  of such systems is a theoretical challenge.  In particular, 
  it seems difficult to produce 
a selective  gravitational collapse which would accrete
 only non-baryonic dark matter without simultaneous baryonic collapse.
  From the
  point of view of dark matter and dark cluster abundance, 
   if  dark clusters include a large fraction of
   mass, then 
   the mass fraction and the cluster abundance in the
universe have been underestimated.
   Whether they contribute significantly
to $\Omega_m$ has to be clarified.  Assuming a flat universe with
  $\Omega_{\Lambda} \approx 0.7$, there is still enough uncertainty
 in $\Omega_m$ to allow dark clusters
 (DC) to contribute up to $\Omega_{m-DC} \approx 0.1$ without
  facing contradiction with what we learned from other mass estimates.

\section{Cosmic shear}
The light propagation through an inhomogeneous universe accumulates 
  weak gravitational deflections over Gigaparsec distances.
   The theoretical works pioneered about 
  40 years ago initiated a long series of theoretical and observational
analysis, which came to successful 
detections two years ago.
   The developments in this field went incredibly fast 
  and revealed some of the interesting cosmological applications of cosmic 
  shear.

\subsection{Gravitational deflection in inhomogeneous universe}
Assuming structures formed from gravitational growth of 
    Gaussian fluctuations, gravitational deflections 
  on cosmological scales can be predicted from Perturbation 
  Theory.  To first order, the convergence $\kappa(\thetag)$ at angular
position $\thetag$ is given by the line-of-sight integral
\begin{equation}
\kappa(\thetag)={3 \over 2} \Omega_0 \ \int_0^{z_s} n(z_s) \ {\rm d}z_s
\
\int_0^{\chi(s)} {D\left(z,z_s\right) D\left(z\right) \over
D\left(z_s\right)} \ \delta\left(\chi,\thetag\right) \
\left[1+z\left(\chi\right)\right] \ {\rm d}\chi \ ,
\end{equation}
where $\chi(z)$ is the radial distance out to redshift $z$, $D$ the angular diameter
distances, $n(z_s)$ is the redshift
distribution of the sources
  and  $\delta$
 the mass density contrast responsible for the deflection at redshift
 $z$.  \  $\delta$ depends on the properties  of the
 power spectrum and tells us how the 
  gravitational convergence field depends on the cosmic 
  history of structure formation.
  Similarly, the cumulative weak lensing effects of
structures also induce a shear field
correlated with the projected mass density which 
   can be characterized
 by the 2-point shear ($ie$ ellipticity) correlation function:
\begin{equation}
\langle\gamma\gamma\rangle_\theta={1\over 2\pi}
\int_0^\infty~P_\kappa(k) J_0(k\theta) \ {\rm d} k,
\label{theogg}
\end{equation}
where $P_\kappa(k)$ is the power spectrum of the convergence field.
Note that to first order we have $\langle\gamma\gamma\rangle_\theta=
\langle\kappa\kappa\rangle_\theta$.
A measurement of this correlation function and of the skewness
of the convergence, $s_3(\theta)$, which probes non Gaussian 
features in the projected mass density field, 
describe most of cosmological 
properties of the convergence field (see \cite{bern97};
\cite{jain97} and references therein). This is easy to demonstrate
with a simple calculation based on perturbation theory and assuming a power
law mass power spectrum. Assuming no cosmological constant, and
a background population at a single redshift
$z$, $<\kappa(\theta)^2>$ and $s_3(\theta)$ can be analytically calculated:
\begin{equation}
<\kappa(\theta)^2>^{1/2} =<\gamma(\theta)^2>^{1/2} \approx 1\% \
 \sigma_8 \ \Omega_m^{0.75} \
z_s^{0.8
} \left({\theta \over  1 {\rm arcmin}}\right)^{-\left({n+2 \over 2}\right)}  \ , \
{\rm and
} \
\end{equation}
\begin{equation}
s_3(\theta)={\langle\kappa^3\rangle\over \langle\kappa^2\rangle^2} \approx 40 \   \Omega_m^{-0.8} \ z_s^{-1.35}  \ ,
\end{equation}
 where $n$ is the spectral index of
the power
spectrum of density fluctuations. It shows that in principle the    
      degeneracy between $\Omega_m$ and $\sigma_8$ is broken 
   when both the variance and the skewness of the convergence 
  are measured. Note that the true relationship between all the cosmological
parameters and the measurement for realistic models (that is including
CDM power spectrum, broad
redshift distribution and non-linear effects) is in fact more complicated, but
these equations faithfully reflect the main dependencies.
\subsection{Expectations}
From an observational point of view, cosmic shear surveys 
  turn out to be a difficult task.
 \cite{vwal99}
   clarified the strategy by exploring  
  the properties of the variance and the skewness of the convergence 
    for various cosmological scenarios.
  They conclude that  the variance can  provide cosmological
 information, provided the survey size is about 1 $deg^2$,
  whereas for the skewness at least 10 $deg^2$ must be covered and 
   more than 100 $deg^2$ for  information on $\Omega_{\Lambda}$
  or  the shape
  of the power spectrum over scales larger than 1 degree.
\\
Like in the precedent Section, assuming perturbation theory and
a power law power spectrum with $n=1$,
we can derive the shear variance as a function of the survey characteristics.
For example for $\Omega_m=0.3$ and $\sigma_8=1$, we have:
\begin{equation}
<\gamma(\theta)^2>^{1/2} = 0.3\%  \ \left[{A_T \over 100
\ deg^2}\right]^{{1 \over 4}} \times \left[{\sigma_{\epsilon_{gal}}
\over 0.4} \right] \times
 \left[{n \over 20 \ arcmin^{-2}}\right]^{-{1 \over 2}} \times
 \left[{\theta \over 10'}\right]^{{-{1 \over 2}}} \ ,
\end{equation}
where $A_T$ is the total sky coverage of the
 survey. The numbers given in the brackets correspond to a measurement
at $3-\sigma$ confidence level of the shear variance. However,
   technical issues regarding corrections of atmospheric and
  optical distortions are the major concerns and 
    the main limitations. The present-day systematic residuals 
   prevent us to measure gravitational shear amplitude smaller than 
  0.3-0.5\% (\cite{erben01}).  Translated into angular scale, it means that 
  we are presently limited to $\theta \approx 2^o$, that is where the expected
signal equals the residual systematics.
\subsection{Observational results}
\begin{figure}
\centerline{\psfig{file=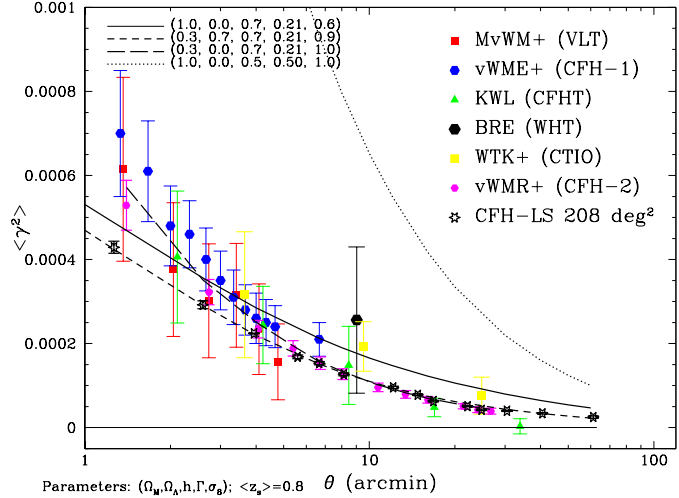,width=7cm,angle=270} }
\caption{Top hat variance of shear as function of angular scale from
  6 cosmic shear surveys. The CFH-LS (open black stars) illustrates 
 the expected signal from a large survey covering 200 deg$^2$. For most
 points the errors are smaller than the stars.
}
\label{sheartop}
\end{figure}

Table \ref{tabcs} lists cosmic shear surveys with already 
  published results. This is not an exhaustive  summary 
  because many surveys are still going on at SUBARU, CTIO, NOAO,
  CFHT, HST or WHT and several are also planned within the next decade
  with new facilities.
  The different strategies adopted by each group enables 
   to handle carefully and in different manner all sources of 
  noise as well as systematics. It is indeed important to
keep
   a variety of approaches in order to cross-check results and
   consistency of cosmological interpretations and, in the future 
  to use all these samples together.
\\
\begin{table}
\begin{center}
{\small
\caption{Present status of cosmic shear surveys with published results.
}
\label{tabcs}
\begin{tabular}{lcccl}\hline
Telescope& Pointings & Total Area & Lim. Mag. & Ref.. \\
\hline
CFHT & 5 $\times$ 30' $\times$30'& 1.7 deg$^2$ & I=24. &
\cite{vwal00}[vWM
E+]\\
CTIO & 3 $\times$ 40' $\times$40'& 1.5 deg$^2$ & R=26. &
\cite{wit00a}[WTK
+]\\
WHT & 14 $\times$ 8' $\times$15'& 0.5 deg$^2$ & R=24. &
\cite{bacon00}[BRE
]\\
CFHT & 6 $\times$ 30' $\times$30'& 1.0 deg$^2$ & I=24. &
\cite{kais00}[KWL
]\\
VLT/UT1 & 50 $\times$ 7' $\times$7'& 0.6 deg$^2$ & I=24. &
\cite{mao01}[Mv
WM+]\\
HST/WFPC2 & 1 $\times$ 4' $\times$42'& 0.05 deg$^2$ & I=27. &
\cite{rhodes01}\\
CFHT & 4 $\times$ 120' $\times$120'& 6.5 deg$^2$ & I=24.
&\cite{vwal01}[vW
MR+]\\
HST/STIS & 121 $\times$ 1' $\times$1'& 0.05 deg$^2$ & V$\approx 26$
& \cite{hammerle01}\\
CFHT & 10 $\times$ 126' $\times$140'& 16. deg$^2$ & R=23.5&
\cite{hoek01a}
\\
CFHT & 4 $\times$ 120' $\times$120'& 8.5 deg$^2$ & I=24.&
\cite{pen01}\\
\hline
\end{tabular}
}
\end{center}
\end{table}
From these surveys, it has been possible to recover the
  amplitude of the cosmic shear variance as function of angular scale.  
  Figure \ref{sheartop} show the remarkable agreement
 between 
 surveys\footnote{ \cite{hoek01a} data are missing because depth is 
  different so the
sources are at lower redshift and the amplitude of the shear
  is not directly comparable to other data plotted}.  This plot is the
most convincing result  showing  the existence of a  correlation
of ellipticities of galaxies in the universe.  It is indeed 
  interpreted as a cosmological weak lensing effect of large-scale
  structures of the universe. However, we should keep in mind that a robust
and definitive cosmological interpretation of these measurements remain dependent
on our ability to pin down the residual systematics and/or to separate signal
and systematics as the E-B mode decomposition seems to be able to make it
(\cite{pen01}). Another important aspect is the necessity to improve our
knowledge of the redshift distribution, which seems possible with
the photometric redshift technique.
\subsection{Cosmological outcomes}
\begin{figure}
\centerline{\psfig{file=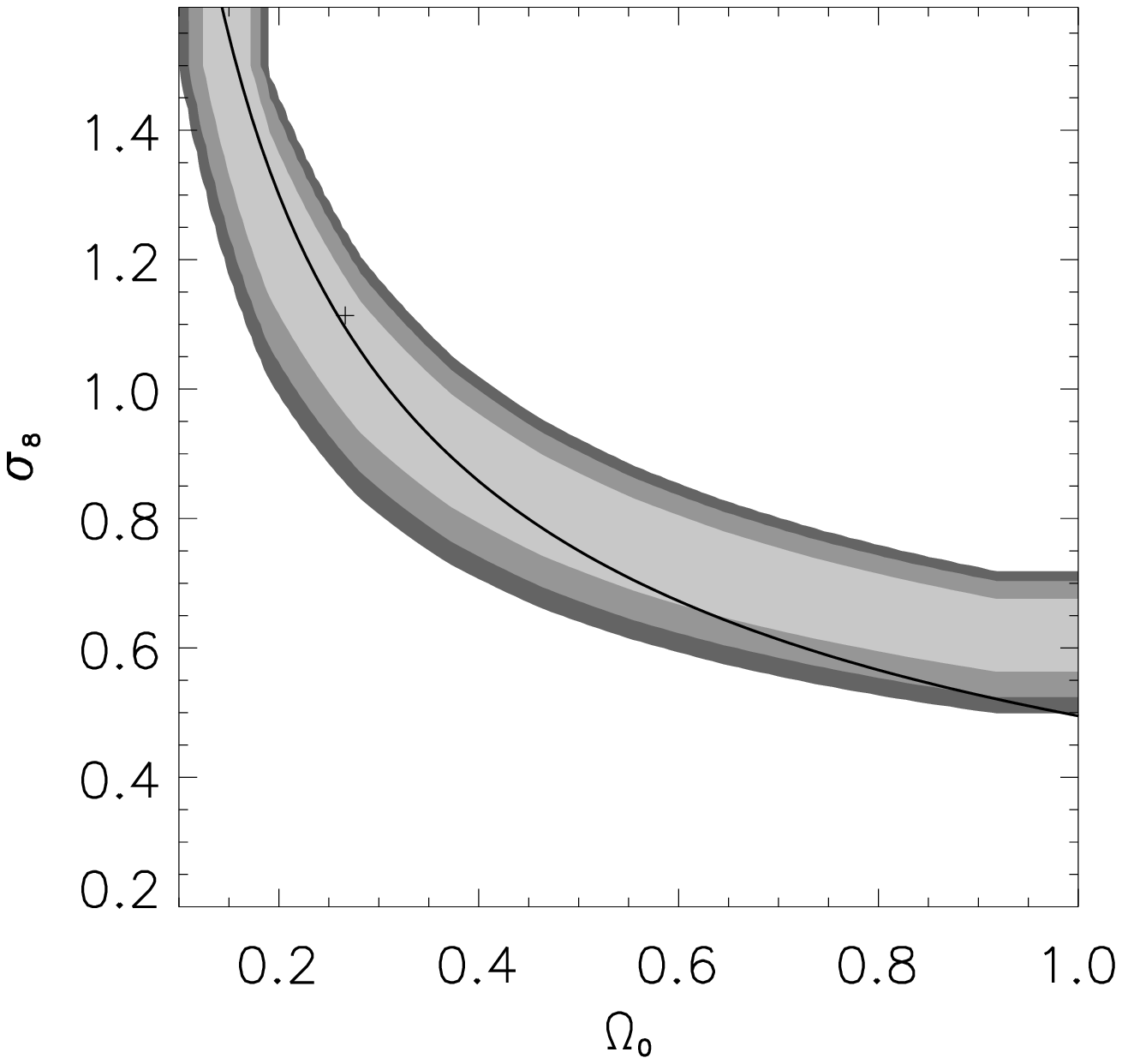,width=6cm}
\psfig{file=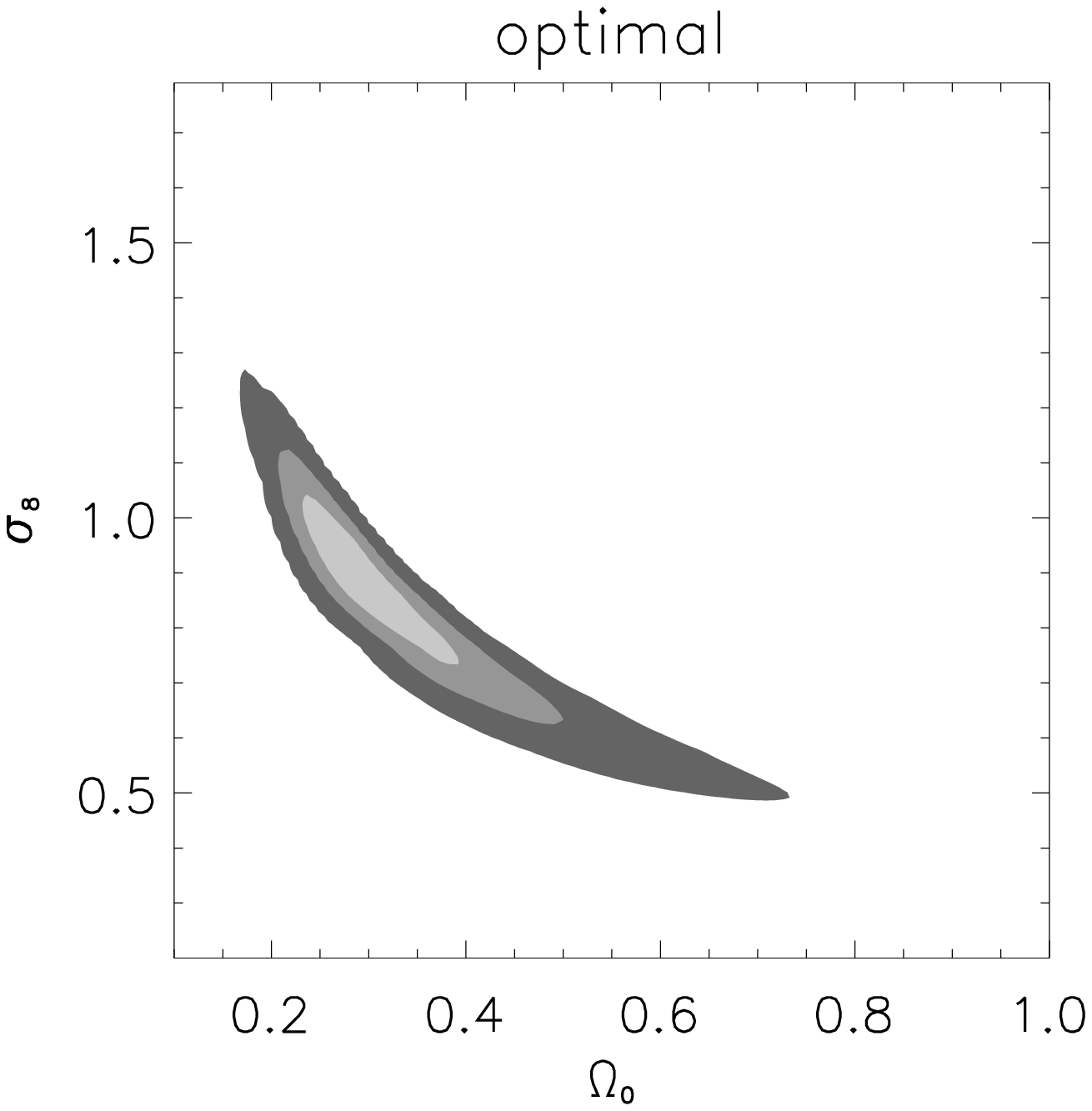,width=6cm}
\psfig{file=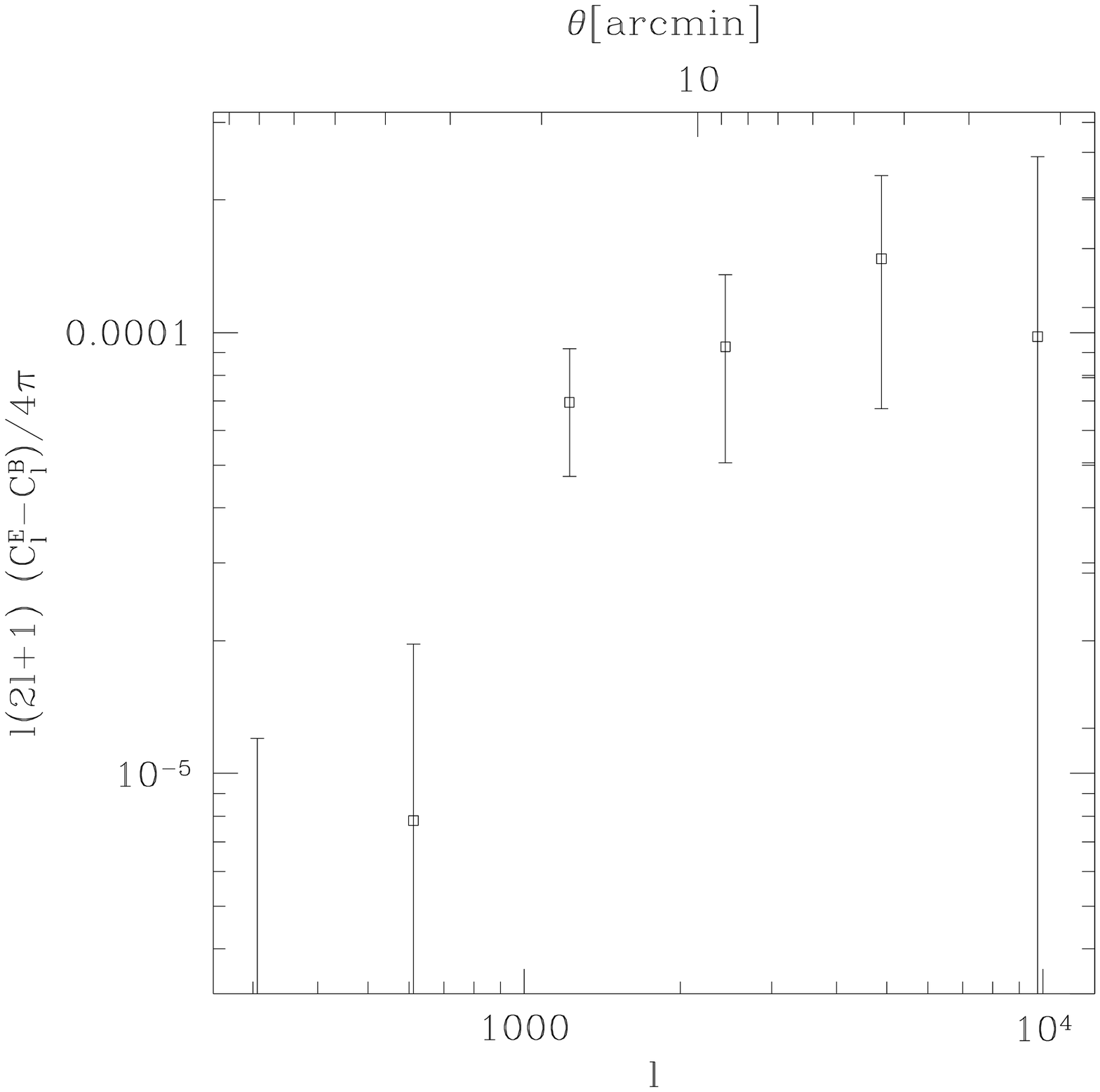,width=4.5cm}
}
\caption{Cosmological results from cosmic shear surveys: 
  the two left-hand panel are 
  $\Omega_m-\sigma_8$ $cl$ contour plots from two independent
data sets.
The left panel is a compilation of five surveys
 (from Maoli et al. 2001). The central panel only
  comes from van Waerbeke et al. (2001).    On the right, 
 the angular power spectrum of the convergence field
  from the VIRMOS-DESCART  survey is plotted (From Pen et al 2001).}
\label{cl}
\end{figure}

Comparison with some realistic cosmological models
  are ploted in Figure \ref{sheartop}. The amplitude of the shear 
  has been scaled assuming that sources are at $<z> \approx 1$, 
   as expected from the comparison of 
  the photometric  depth of these surveys with the deepest spectroscopic 
surveys done so far.  The
 standard COBE-normalized CDM is ruled  at a 10$-\sigma$  confidence
level.  However, there are  many models which fit the data.
 This simply illustrates the degeneracy between $\Omega_m$ and 
   $\sigma_8$ we discussed in the previous section and 
 which   cannot be broken without high-order statistics.  
  A plot with  
 $\Omega_m$-$\sigma_8$ confidence level contours is more 
   meaningful, as in Fig. \ref{cl}.
   The left and middle
panels show two independent data sets, both containing almost the
same number of galaxies. The left panel is a compilation 
  of the five first survey of the previous figure.  It covers 
  6.5 deg$^2$ over  75 independent areas and contains 
   250,000 galaxies.  The middle panel is the CFH-2 survey 
  (see figure \ref{sheartop}) which covers
 8.5 deg$^2$ over 4 independent fields and contains 
   450,000 galaxies.   Both $\Omega_m$-$\sigma_8$ contours
  are difficult  to reconcile observations with an
$\Omega_m>0.8$-universe. Assuming a CDM model with 
  $\Gamma \approx 0.2$, we can
conclude that reasonable cosmological models have:
\begin{equation}
0.05 \le \Omega_m \le 0.8 \ \ \ \ {\rm and} \ \ \ \ 0.5 \le \sigma_8 \le
1.2 \
.
\end{equation}
A more detailed investigation of the data also permits to 
   probe the power spectrum of the dark matter on scale below
  30 arc-minutes. 
    \cite{pen01}  cleaned first  the CFH-2 sample of the 
  VIRMOS-DESCART
  cosmic shear survey\footnote{http://terapix.iap.fr/Descart and 
 http://www.astrsp-mrs.fr/virmos/} 
   from systematics residuals. This is possible by separating
the shear field into curl-free and curl modes (called
E and B modes respectively). This is identical to the
technique expected to take place on the cosmic microwave
background polarization analysis. In the case of weak lensing, the
residual are expected to contribute equally to the E and B modes, while
the cosmic shear signal shows up only in the E channel.
    By  separating the $E-$ and $B-$ components of the shear 
   field, \cite{pen01} succeeded to reduce errors by a factor of $\sqrt{2}$ 
   of the pure gravitational lensing signal.  Then, 
  they computed each $C(l)$ from the $E$-mode in a standard way.
 They are shown on
Figure \ref{cl}.  Since they are  inferred
  from weak distortion of galaxies only,  they represent 
    a direct measurement the projected dark matter
 power spectrum.  The data cover a small area,
    are still confined inside
 the range $1' \le \theta \le 30'$ ($5 \times 10^2 \le l \le 10^4$) and
are
noisy.  However, the signal is strong enough to be compared with 
   cosmological models or other $C(l)$ inferred from 
   redshift surveys or the CMB.  It shows the 
  potential of future cosmic shear surveys, in particular that it will be possible
to measure the 3D mass power spectrum with accurate measurements.
\subsection{Systematics and calibration issues}
There are four major concerns regarding the amplitude and the 
interpretation of the weak distortion signal.  \ \ \ (1) systematics 
produced by wrong or insufficient PSF anisotropy correction; 
  \ \  \ (2) systematics produced by real but non-lensing signal 
  producing similar distortion as cosmic shear, like 
intrinsic correlations of ellipticities \ \ \ (3) 
  wrong scaling of redshift distribution of sources, and 
 \ \ \ (4) source clustering.   
\\
All of them have been analyzed at length.
   Careful tests regarding PSF anisotropy  corrections
 demonstrate  that residuals are
 small on scale ranging from 1 arc-minute to 30
arc-minutes.  The intrinsic correlations of ellipticities
 which could be generated during galaxy formation processes
  may produce similar signatures as cosmic shear.  Several
  recent  numerical and theoretical  studies (see for
 example \cite{crit00}; \cite{mackey01})
 show that intrinsic correlations are negligible on scales beyond
  one arc-minute, provided that the survey is deep enough. In that case,
  most lensed galaxies along a line of sight are spread over
 Gigaparsec scales and have no physical relation with its apparent 
  neighbors.
  Since most cosmic survey are deep, they are almost
    free of intrinsic correlations.
     \cite{pen01}, in particular,
    have confirmed  that the VIRMOS-DESCART
   survey listed in Table \ref{tabcs}
  shows  a pure cosmic shear signal on scales beyond
one arc-minute, which dominates any systematics  by
    significant factor.
  In contrast, shallow surveys may have strong contamination. Although
 it weakens the interest of shallow survey for cosmic shear, they
are nevertheless interesting since 
 it give us insight on  properties of intrinsic ellipticity correlations
  produced by the generation of angular momentum during merging 
processes of halos.
\\
Likewise redshift of arcs which scale the total mass inside a 
critical radius,  the redshift of sources used for cosmic shear 
analysis scales the amplitude of the shear.  Current surveys 
   probe sources up to $I \approx 24$.  This is within the 
   limits of giant telescopes, so
detailed informations on redshift distributions of lensed galaxies
  will be soon well constrained\footnote{Using the DEEP2 (\cite{dav00}
 Davis et al. 2000)
 or VIRMOS (\cite{olf00} Le F\`evre et al. 2000) surveys for example.}. Beyond this limit, there 
 is no possibility to get spectroscopic redshifts of sources and 
  only photometric redshifts are expected to work.  Present-day 
calibrations show that the technique works well enough to provide
  source redshift with sufficient accuracy up to $I \approx 25$. 
  Beyond this, it may be more critical.
\\
The difficulties may be with clustering of sources which seems 
  to significantly affect high-order statistics, like skewness
 (\cite{hamana00}). Uncertainties on the amplitude of 
  the skewness  produced by clustering may be as high as 30\% . 
If so, multi-lens plane cosmic shear analysis will be necessary
  which implies a good knowledge of the redshift distribution.
   For very deep cosmic shear surveys, this could be  could be 
  a challenging issue.
\section{Conclusions and prospects}
Tracing the matter and its evolution with look-back time
    is a major goal of all massive 
surveys done with ground based telescopes or satellites.  
  Weak gravitational lensing is a new but reliable tool for this 
purpose because it is almost insensitive on the nature 
  and the physical stage of the matter. Past and present
    experiences show that 
   it can provide astrophysical informations at about all scales
  ranging from 10 kpc to 1 Gpc and address as well key scientific question 
  relevant for fundamental physics.
\\
In this review we focussed on galaxies, cluster of galaxies and 
large-scale structures. But first tentatives by Hoekstra at al 
 (\cite{hoek01b}) on the CNOC2 surveys on groups of galaxies and by Kaiser et 
al 1998 (\cite{kais98}) 
  on a supercluster of galaxies will be soon widely deployed on very 
large samples in order to understand the physics of these systems
  and their interactions with galaxy halos, clusters and very 
large structures. Used jointly with cosmic shear analysis, we 
  expect that  these 
surveys will provide a complete description of the dark matter, with 
particular emphasis on the properties of its power spectrum,  the 
radial mass profile of bound systems and the coupling between 
  baryonic and non-baryonic matter.  This last point is now underway.
  The recent analysis on the biasing and galaxy-mass cross correlation
 coefficient in the RCS survey by Hoestra et al show that we can 
now test whether the linear biasing is valid and probe the relation 
between mass and light as function of angular 
 scale (\cite{schn98}; \cite{vwa98}).  Similar studies can be done inside clusters of 
  galaxies by using jointly weak lensing, X-ray and SZ reconstructions
  (\cite{zar97}, \cite{dore01}).
\\
 Next cosmic shear survey generation with MEGACAM at CFHT or VST/VISTA
 at Paranal   
 or even space based panoramic cameras will improve by one  order of
 magnitude in details and precision (see figure \ref{sheartop}).
  It is expected that they will provide accurate projected mass 
  reconstruction, similar to APM galaxy survey. They should be able   
to break the degeneracy between $\Omega_m$ and $\sigma_8$  from
  the analysis of the skewness of $\kappa$. On  a longer
  time-scale, very large cosmic shear surveys 
   will probe the dark matter power spectrum over scales 
  larger than 10 degrees and will permit to  constrains  $\Omega_{\Lambda}$,
 or any quintessence fields (\cite{wu99}; \cite{benab01}).

\section*{Acknowledgements}  
We thank R. Athreya, K. Benabed, F. Bernardeau, E. Bertin, O. Dor\'e,   
 B. Fort, H. Hoekstra, B. Jain, P. Schneider, for useful 
   discussions.  This work was supported by the TMR Network ``Gravitational 
 Lensing: New Constraints on
Cosmology and the Distribution of Dark Matter'' of the EC under contract
No. ERBFMRX-CT97-0172. YM thanks the organizers of the meeting for 
  financial support.

\vfill
\end{document}